\documentclass[%
reprint,amsmath,amssymb,aps]{revtex4-1}

\usepackage{graphicx}
\usepackage{dcolumn}
\usepackage{bm}
\usepackage[T1]{fontenc}
\usepackage{mathptmx}
\usepackage{hyperref}

\usepackage{color}
 
\usepackage{xr}
\makeatletter
\newcommand*{\addFileDependency}[1]{
  \typeout{(#1)}
  \@addtofilelist{#1}
  \IfFileExists{#1}{}{\typeout{No file #1.}}
}
\makeatother

\newcommand*{\myexternaldocument}[1]{
    \externaldocument{#1}
    \addFileDependency{#1.tex}
    \addFileDependency{#1.aux}
}

\myexternaldocument{Spatial_modulation_sm}
\begin{document}


\title{Spatial modulation of nonlinear flexural vibrations of membrane resonators}

\author{Fan Yang}
\author{Felix Rochau}
\author{Jana Huber}
\author{Alexandre Brieussel}
\author{Gianluca Rastelli}
\author{Eva M. Weig}
\author{Elke Scheer}
 \email{elke.scheer@uni-konstanz.de}

\affiliation{Department of Physics, University of Konstanz, 78457 Konstanz, Germany}

\date{\today}

\begin{abstract}
We study the vibrational motion of membrane resonators upon strong drive in the strongly nonlinear regime.
By imaging the vibrational state of rectangular siliconnitride membrane resonators and by analyzing the frequency response using optical interferometry, 
we show that upon increasing the driving strength, the membrane adopts a peculiar deflection pattern formed by concentric rings superimposed onto the 
drum head shape of the fundamental mode.
Such a circular symmetry cannot be described as a superposition of a small number of excited linear eigenmodes.
%
%
Furthermore, the different parts of the membrane oscillate at different multiples of the drive frequency, an observation that we denominate as  'localization of overtones'. 
We introduce a phenomenological model that is based on the coupling of a very small number of effective nonlinear oscillators, representing the different parts of the membrane, and 
that describes the experimental observations. 
%
%
%
\end{abstract}

%

\maketitle

%
%
Micro- and nanoscale mechanical resonators bear rich potential to expand their applications in a 
broad range of areas, such as 
noise sensing \cite{weber2016force}, 
optical quantum metrology \cite{kampel2017improving,aspelmeyer2014cavity}, 
nanoelectromechanical logic gates \cite{mahboob2008bit, guerra2010noise} 
and parametric oscillators \cite{rugar1991mechanical, seitner2017parametric}. 
Mechanical resonators can have different size scales, and may be even as small as single carbon nanotubes or patterned graphene sheets \cite{eichler2011nonlinear, zhang2015vibrational}. 
Among larger devices, quasi-two dimensional membrane oscillators are interesting for several reasons, e.g. they have resonant behavior in a broad range of frequencies (e.g. from 100 kHz to several MHz) which make them attractive as broad-band transducers for vibratory energy harvesting \cite{jia2016twenty}. 
They are also important in quantum engineered systems, since flexural modes can be easily coupled to other degrees of freedom  such as photons \cite{sankey2010strong,purdy2013observation,andrews2014bidirectional,xu2016topological} and cold atoms in hybrid optomechanical architectures \cite{camerer2011realization,jockel2015sympathetic}.
Suspended membrane resonators operating in the single-mode regime have been studied in the weakly nonlinear limit and modeled as a Duffing oscillator \cite{antoni2013nonlinear}. 
Increasing the membrane oscillation amplitudes enforces the nonlinearity, thereby giving rise to nonlinear coupling between eigenmodes \cite{nayfeh2008nonlinear,manevich2005mechanics,schuster2009reviews}.
So far, experiments reported nonlinear interaction involving only very two interacting modes \cite{van2010amplitude,westra2010nonlinear,antonio2012frequency,guttinger2017energy,chen2017direct}, and only when matching an internal resonant condition between eigenfrequencies \cite{shoshani2017anomalous}. 
%
%
%
%
%
%

In this Letter we analyze a novel, nonlinear vibrational state of a membrane resonator, achieved under strong drive.
We observe the appearance of spatially modulated overtones. 
Upon increasing the driving strength, the membrane adopts a characteristic ring-shaped deflection pattern in which different parts of the membrane oscillate at different subharmonically excited multiples of the  driving frequency.
Such a ring-shaped pattern is distinctly different from the waveform and the symmetry of the flexural eigenmodes of the rectangular membrane.   
In other words, this regular pattern can only be reconstructed by using a superposition of a multitude of eigenmodes of the membrane. 
Moreover, the presence of subharmonically excited tones points out that the system is in a strongly nonlinear vibrational regime. 
Part of this phenomenon was reported before \cite{zhou2017non}, but remained unexplained.
Here we analyze it in detail and  present a phenomenological model that captures the main features and semi-quantitatively describes the experimental findings. 
The starting point of this model is that, under driving, the profile is well described by
\begin{equation}
\label{eq01}
u(r,t) \cong 
\sum_{n\geq 1} q_n(t) h_n\left(r\right) e^{i 2n\pi f_\mathrm{d}t}  
+
\mbox{c.c.}
\, ,
\end{equation}
where $r$ is the radial coordinate of the 2D membrane, $\textit{h}_n(r)$ 
are arbitrary spatial profile functions, and the amplitudes of oscillation 
are $\textit{q}_n(t)$, respectively.
The amplitudes $\textit{q}_n(t)$ play the role 
of effective oscillators with different eigenfrequencies $f_n$ and having sizeable amplitudes at different positions on the membrane, described by the functions $\textit{h}_n(r)$, i.e. the localized overtones.
 We assume a nonlinear interaction $V = \sum_{n\geq 2} \lambda_n q_1^n q_n$ between a mode $q_1$ with eigenfrequency $f_1$ and its overtones. This model is referred to as $1:n$ nonlinear coupling model.
%
%
%
%
%

We will show that, to describe our experimental observations, a small number of modes is sufficient, representing a significant simplification compared to the classical model based on linear eigenmodes of a rectangular membrane.  
Our conclusions are thus that spatially modulated overtones represent, indeed, a new paradigm in the nonlinear dynamics of membranes 
which goes beyond previous theoretical pictures, based on the approach of nonlinear interaction of few linear eigenmodes \cite{antonio2012frequency,guttinger2017energy,chen2017direct}.  

\begin{figure}[t]
  \includegraphics[width=\linewidth]{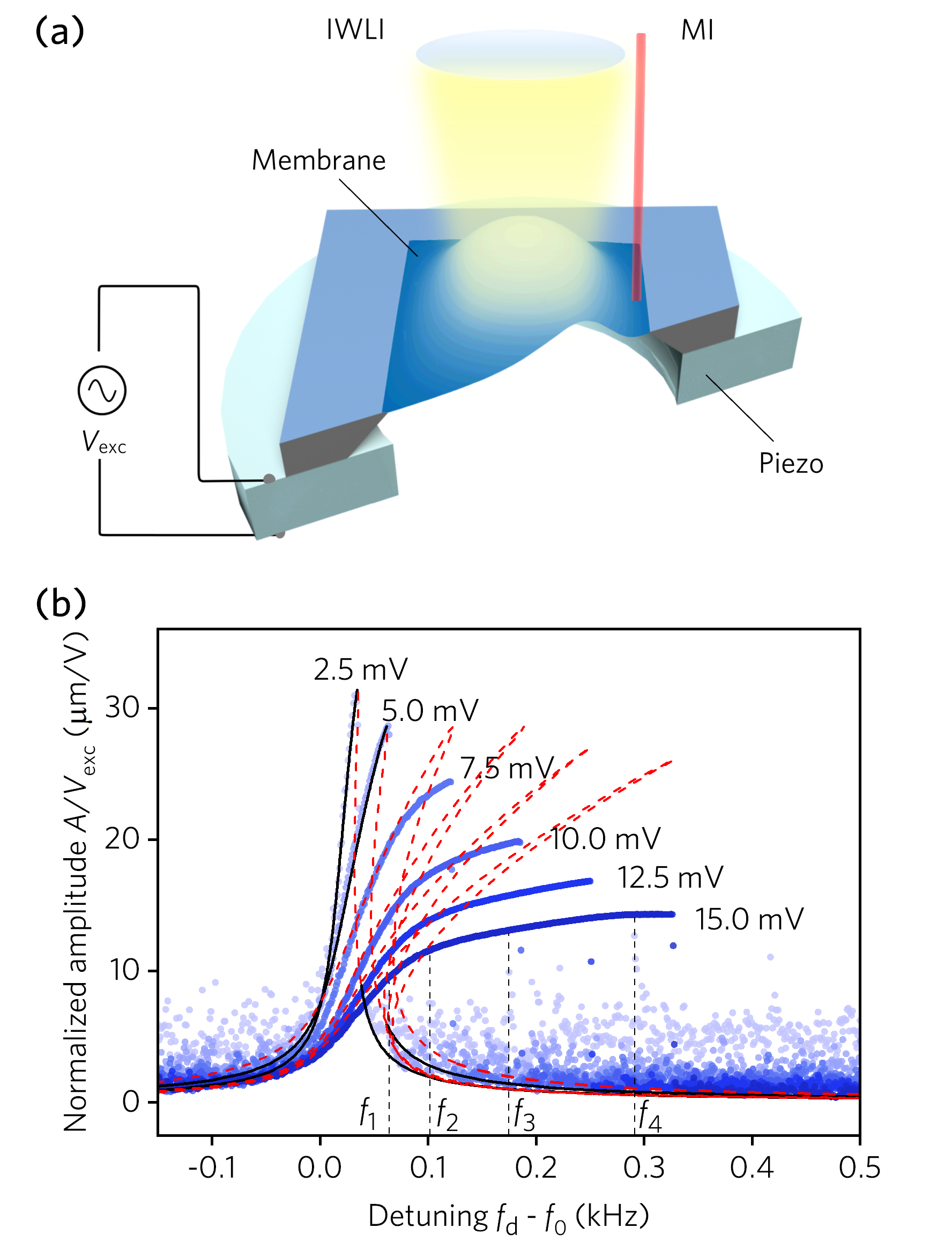}
  \caption{(a) Sketch of the experimental setups showing the membrane chip carrying a free-standing SiN membrane, the piezo ring and drive, the objective of the imaging white light interferometer (IWLI) and the laser beam of the Michelson interferometer (MI). (b) Nonlinear resonance curves with varying $\textit{V}_{\textrm{exc}}$, normalized to $\textit{V}_{\textrm{exc}}$, are plotted vs the detuning frequency with respect to the linear eigenfrequency $\textit{f}_{\textrm{d}}-\textit{f}_0$. The red dashed lines are fits with a Duffing model.}
  \label{fig:image01}
\end{figure}

%
%
%
The sample fabrication and measurement principles of the siliconnitride (SiN) membranes have been described in detail 
elsewhere \cite{petitgrand20013d, hernandez2002photoacoustic, waitz2012mode, waitz2015spatially, yang2017quantitative} 
and are summarized in the Supplementary Material (SM). 
%
Flexural modes of the membrane are excited by applying an AC voltage 
$V_\textrm{exc}$ $\cdot$ sin(2$\pi \textit{f}_\textrm{d}$\textit{t}) to the piezo ring that causes a uniform thickness change of the piezo, 
resulting in an inertial excitation of the membrane, see Fig.~\ref{fig:image01}(a). 
The vibrational state of the membrane installed in a vacuum chamber is observed by two different types of optical interferometry. 
The imaging white light interferometer (IWLI) is able to spatially resolve the deflection profile and 
to obtain the average amplitude response by integrating the deflection profile over a selected area 
on the membrane surface \cite{petitgrand20013d}. 
The Michelson interferometer (MI) is focusing on one particular position of the membrane with 
a spot diameter of $\mathrm{\sim}$2 $\mu$m. 
Further experimental details are given in the SM.\\
%
%
%
%
%
%
%
In the linear response regime, we find resonances corresponding to the eigenfrequencies $\textit{f}_{(\textit{m,n})}$ 
of the flexural eigenmodes which are characterized by the integers \textit{m} and \textit{n} indicating the number 
of deflection maxima in the two spatial directions of the membrane plane \cite{yang2017quantitative}. 
For example, we find $\textit{f}_{\textrm{(1,1)}}$ = 321 kHz and $\textit{f}_{\textrm{(2,2)}}$ = 646 kHz for the eigenfrequencies of (1,1) and (2,2) 
mode, respectively, both with quality factors in the order of 20000. 
%
%
%
The eigenfrequencies of all modes discussed in this article as well as further mechanical parameters are listed in Table 1 in the SM.\\
%

\begin{figure*}[t]
  \includegraphics[width=\textwidth]{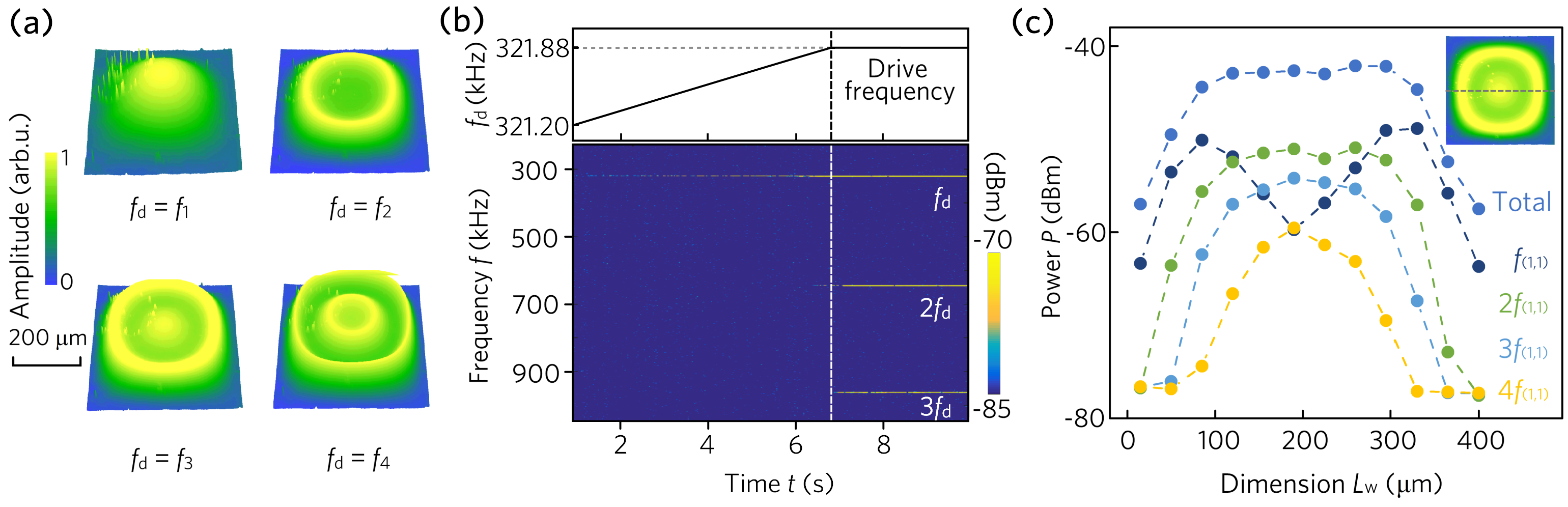}
  \caption[width=\textwidth]{ Localized overtones for near-resonant drive in the spatial modulation regime. (a) Spatial deflection patterns observed at different driving frequencies $\textit{f}_{\textrm{d}}$ denoted as $\textit{f}_{\textrm{1}}$, $\textit{f}_{\textrm{2}}$, $\textit{f}_{\textrm{3}}$, $\textit{f}_{\textrm{4}}$ in Fig. \ref{fig:image01}(b), associated with the spatial overtones of the (1,1) mode ($\textit{V}_{\textrm{exc}}$ = 16.2 mV). (b) Power spectrum (measured by MI) of a sweep-up measurement around the (1,1) mode up to $\textit{f}_{\textrm{d}}$ = 321.88 kHz (dashed vertical white line), $\textit{V}_{\textrm{exc}}$ = 50 mV. Top: drive frequency  as a function of time. (c) Position dependence of relative amplitude of overtones recorded at $\textit{V}_{\textrm{exc}}$ = 35 mV and $\textit{f}_{\textrm{d}}$ $\approx$ $\textit{f}_{\textrm{3}}$ (see Fig. \ref{fig:image02}(a) and measured by MI) for a line cut parallel to the edges through the center of the membrane, see schematic in the inset. The intensities of the individual overtones (up to 4$^{\textrm{th}}$ overtone shown) and the total intensity are obtained from the sweep-up measurements. }
  \label{fig:image02}
\end{figure*}

We utilize the IWLI signal integrated over the entire membrane area to record the nonlinear vibration behavior 
under intermediate and strong sinusoidal excitation and with a driving 
frequency $\textit{f}_{\textrm{d}}$ around $\textit{f}_{\textrm{(1,1)}}$. 
With $\textit{V}_{\textrm{exc}}$ = 2.5 mV to 5.0 mV, the resonance curves exhibit a Duffing-type nonlinearity, 
as shown in Fig.~\ref{fig:image01}(b), but when exceeding $\textit{V}_{\textrm{exc}} \mathrm{>}$ 5.0 mV, 
the amplitude rises weaker than expected for the Duffing model but persists over a larger frequency range. 
This observation signals the onset of the spatial modulation phenomenon. 
The spatial deflection profiles captured by IWLI for $\textit{V}_{\textrm{exc}}$ = 15.0 mV are shown in Fig.~\ref{fig:image02}(a). 
When continuously sweeping the driving frequency $\textit{f}_{\textrm{d}}$, the spatial appearance 
of the deflection pattern changes from single drum head type at $\textit{f}_{\textrm{d}}$ = $\textit{f}_{\textrm{1}}$ 
(see marker in Fig.~\ref{fig:image01}(b)) to a crater type at $\textit{f}_{\textrm{d}}$ = $\textit{f}_{\textrm{2}}$, 
where the amplitude at the membrane center decreases and the central part of the deflection pattern adopts a flat shape. 
When further increasing $\textit{f}_{\textrm{d}}$ to $\textit{f}_{\textrm{3}}$, the flat area evolves into a separate drum head shape. 
Finally, at $\textit{f}_{\textrm{4}}$ again a minimum occurs in the center surrounded by two rings. 
The observed deflection patterns reveal a frequency-dependent spatial modulation of the (1,1) mode, 
which cannot be described as a simple superposition of few (\textit{m,n}) modes, since it has a circular symmetry 
contrary to the axial symmetry of the higher (\textit{m,n}) modes in a rectangular membrane. 
%

%
With the help of sweep-up and ring-down experiments performed in the MI at different positions on the membrane, one can reveal that the modulation is caused by frequencies which are identified as 
overtones of the (1,1) mode with 2$\textit{f}_{\textrm{(1,1)}}$ and 3$\textit{f}_{\textrm{(1,1)}}$, as we will show in the following.
%
%
%
%
We perform a slow sweep-up of $\textit{f}_{\textrm{d}}$ at constant $\textit{V}_{\textrm{exc}}$ 
from a starting value below the resonance frequency of a particular mode up to a maximum frequency above the resonance frequency but within the frequency range with large amplitude. 
At this frequency we switch off the drive. 
The measurement spot is close to one edge of the membrane and chosen such that the maximum deflection stays smaller than a quarter of 
the laser wavelength (see SM for details). 
A Fast Fourier Transform (FFT) algorithm is performed on the time domain oscillation displacement data in narrow 
time windows to get a time-resolved power spectrum of the sweep-up and ring-down processes. 
Figure~\ref{fig:image02}(b) shows the time evolution during the sweep-up of the power spectrum for a moderate 
excitation for $\textit{f}_{\textrm{d}}$ = 321.88 kHz. 
The subsequent ring-down part is displayed in Fig. S\ref{fig:image11} in the SM section 6. 
In contrast to the linear response regime shown in Fig. S\ref{fig:image10} in the SM, we observe response 
at multiple frequencies which are setting in one after another at different $\textit{f}_{\textrm{d}}$. 
Strong signals are observed at $f_\textrm{d}$, $2 f_\textrm{d}$, and $3 f_\textrm{d}$, respectively.
These responses evolve during ring-down towards the overtones of the first mode,
namely to $f_{(1,1)}$, $2 f_{(1,1)}$, and $3 f_{(1,1)}$.
We note that 2$\textit{f}_{\textrm{(1,1)}}$ = 642 kHz can clearly be distinguished from 
the eigenfrequency $\textit{f}_{\textrm{(2,2)}}$ = 646 kHz of the (2,2) mode, given 
the frequency resolution of the FFT algorithm $\mathrm{\sim}$ 1 kHz. 
Similar multiple frequency response has also been found for SiC and Ge membrane systems \cite{zhou2017non}. 
The multiple frequency response complements the observations from the IWLI and gives a strong hint that the spatial modulation state 
corresponds to spatially varying superpositions of the ground mode and its overtones. 
In addition, stroboscopic IWLI measurements show (see Fig.~S\ref{fig:image08}(d) in the SM), 
that at $\textit{f}_{\textrm{d}}$ = $\textit{f}_{\textrm{3}}$ the central part of the membrane (inner disc with a drum head shape) 
oscillates at different frequency than the outer annulus which oscillates at $\textit{f}_{\textrm{d}}$.

To further test this hypothesis of localized overtones we perform MI measurements at different positions on the membrane. 
Figure~\ref{fig:image02}(c) shows the relative amplitudes of the overtones observed in the power spectrum 
for a line profile from one edge to the other, crossing the center of the membrane, see inset of Fig.~\ref{fig:image02}(b). 
At each spot a sweep-up is performed and the intensities of the overtones are evaluated at fixed detuning 
at the various spots by separately integrating the power of each time window within a corresponding frequency range. 
The total intensity reveals a large plateau that extends over a width of roughly 200 $\mu$m  in the center of the membrane. 
The fundamental frequency $f_\mathrm{d}$ has the highest amplitude at the outer rim of this plateau and strongly decreases towards the center. 
In the central part the intensity of the 2$^{\textrm{nd}}$ overtone exceeds the one of the ground mode. 
The 3$^{\textrm{rd}}$ and 4$^{\textrm{th}}$ mode have weaker intensity and are even more centered.
This observation evidences the existence of spatially localized overtones, confirming the interpretation of the IWLI data.
%
%
%
%
%
%

The detailed microscopic analysis of the observed spatial overtones and their appearance is a challenging theoretical task 
and beyond the aim of this work.
%
%
The main physical aspects can be captured by the phenomenological ansatz given in Eq. (\ref{eq01}) that describes that the membrane is vibrating at 
distinct sections at different frequencies.
We note that any profile function can be obtained as a linear combination of the eigenmodes (\textit{m},\textit{n}) 
since the latter form a complete orthonormal basis. 
An example is given in the SM in which we project the ring-shape of the deflection onto the linear eigenmodes.
This yields a very slowly convergent series: as expected intuitively, a superposition of many  modes is needed to reproduce 
a circular ring-shape  (\textit{N} $\mathrm{\ge}$ 100) of the rectangular membrane (see SM).
By contrast, the spatially modulated overtones $\textit{u}_n$ can be viewed as renormalized eigenmodes appearing only for strong driving.
They can be mathematically defined on the concept of an invariant manifold in an elastic nonlinear medium \cite{pesheck2001nonlinear}.
For simplicity we describe here the situation with only two oscillators, i.e. the 1:2 model. The general $1:n$ model is detailed in the SM.
We thus consider the equation
\begin{equation}
\label{eq02}
u(r,t) \cong 
q_\mathrm{2}(t) h_\mathrm{2}\left(r\right) e^{i 4\pi f_\mathrm{d}t}  
+
q_\mathrm{1}(t) h_\mathrm{1}\left(r\right) e^{i 2\pi f_\mathrm{d}t}  
+
\mbox{c.c.}
\, ,
\end{equation}
%
%
%
%
$\textit{h}_{\textrm{1}}(t)$ is localized at a ring of radius \textit{R} from the center and 
has the eigenfrequency $f_{\textrm{1}} \simeq \textit{f}_{\textrm{(1,1)}}$, the second one, and $\textit{h}_{\textrm{2}}(t)$,
is localized at the center, with eigenfrequency $2 f_{\textrm{1}}$, as shown in 
Fig.~\ref{fig:image02}(a) for $\textit{f}_{\textrm{d}}$ = $\textit{f}_{\textrm{3}}$. 

Finally, in our model, we assume a nonlinear coupling between the two effective oscillators and we set the following coupled equations 
for the amplitudes $\textit{q}_\textrm{1}$ and $\textit{q}_\textrm{2}$:
\begin{align}  
{\ddot{q}}_{\mathrm{2}}(t)
& =
- 4{(2\pi f_{\mathrm{1}})}^2 q_{\mathrm{2}}(t) 
- 2{\mathrm{\Gamma }}_{\mathrm{2}}{\dot{q}}_{\mathrm{2}}(t)
- \, \lambda \, (q_{\mathrm{1}})^2 / 2 
\label{eq03}  
\\
{\ddot{q}}_{\mathrm{1}}(t)
& = 
-{\left(2\pi f_{\mathrm{1}}\right)}^2  q_{\mathrm{1}}(t) 
- 2{\mathrm{\Gamma }}_{\mathrm{1}}{\dot{q}}_{\mathrm{1}}(t)
- 2\lambda \, q_{\mathrm{1}}(t) q_{\mathrm{2}}(t) 
\nonumber\\
& \,\, -\gamma_{\mathrm{1}}(q_{\mathrm{1}})^3(t)  + F_{\mathrm{d}}{\mathrm{cos} \left(2\pi f_{\mathrm{d}}t\right)} 
\label{eq04}
\end{align}
Equations \eqref{eq03} and \eqref{eq04} mean that the membrane is effectively composed of an inner disc coupled to an outer ring.  
%
%
The ring oscillator is characterized by a Duffing nonlinearity with strength $\alpha_\textrm{1}$, 
and both are damped with damping constants $\Gamma_\textrm{1}$ and $\Gamma_\textrm{2}$. 
Most importantly, we also postulate a nonlinear resonant interaction between the two oscillators, 
given by the last terms in Eqs. \eqref{eq02} and \eqref{eq03} with the interaction strength $\lambda$, which can be derived from the nonlinear potential $V =\lambda {q_{\mathrm{1}}}^2 q_{\mathrm{2}}/2$.
This interaction is important since it implies a correlated behavior of the two amplitudes which we actually observe in the system, as we discuss next.
%
%
%
%
\begin{figure}[t]
  \includegraphics[width=\linewidth]{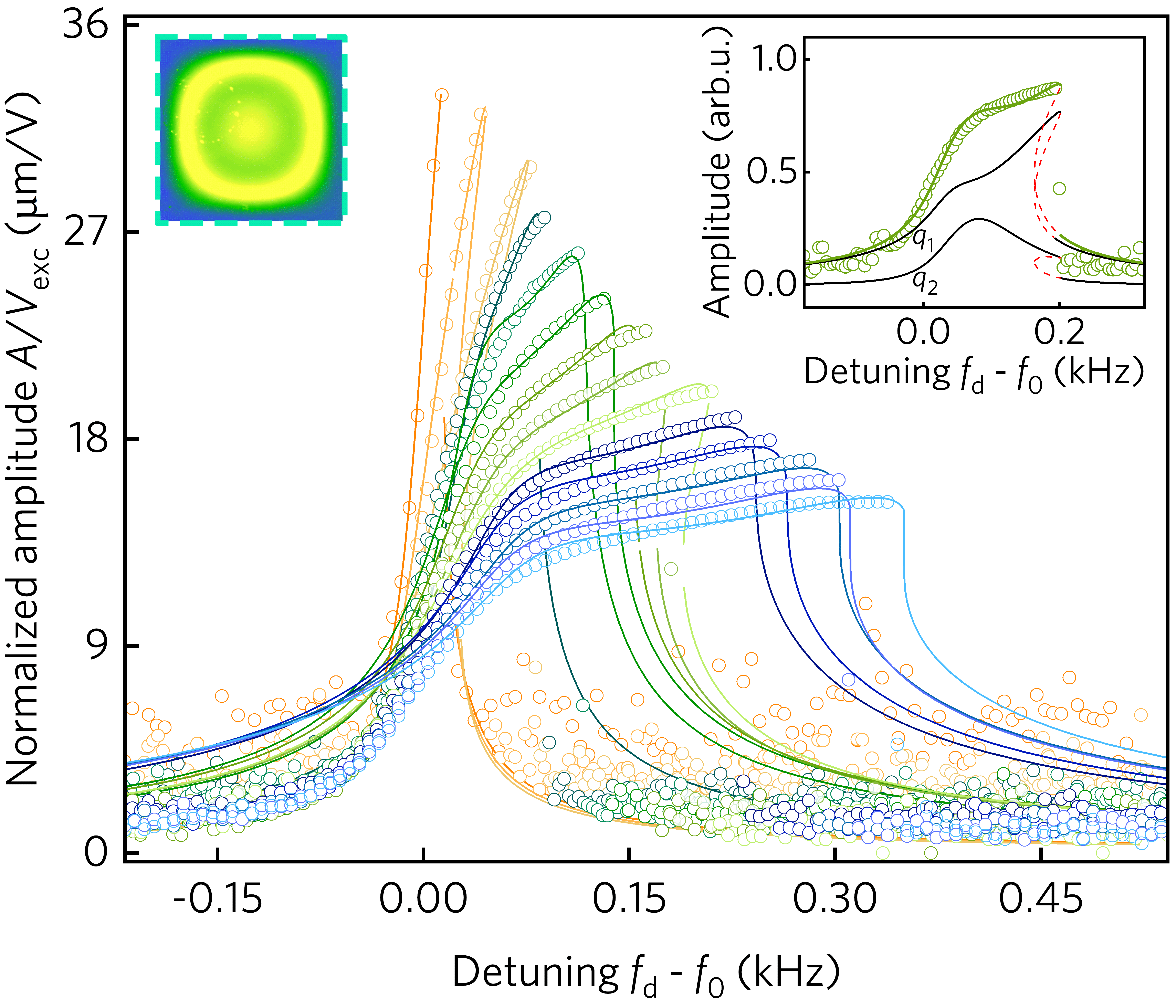}
  \caption{Experimental responses (symbols) of the full membrane area, integrated over the entire membrane (shown as green frame in the inset) measured by IWLI for $\textit{V}_{\textrm{exc}}$ = 2.5, 4.2, 5.8, 6.7, 7.5, 8.3, 9.1, 10.0, 10.8, 11.6, 12.5, 13.3, 14.1, and 15.0 mV at \textit{P} = 0.055 mbar. The lines are fits with either the duffing or the coupled model. For details see text.  Inset: Fitting of the curve recorded at 10.0 mV. The black lines show the individual contributions of modes $n=1$ and $n=2$.}   
  \label{fig:image03}
\end{figure}

In Fig.~\ref{fig:image03} we show more curves from the same data set as depicted in Fig.~\ref{fig:image01}(b). The orange lines are the same fits to the duffing model as in Fig.~\ref{fig:image01}(b), the green curves are calculated with the 1:2 coupling model and the blue curves using  1:2 and 1:3 coupling. The fitting parameters and their analyses are given in the SM.
%
%
The model fits the experimental curves well. 
The small deviations can be explained 
by the simplicity of the model, e.g. neglecting the contributions of the higher overtones. 
%
%

\begin{figure}[htbp]
  \includegraphics[width=\linewidth]{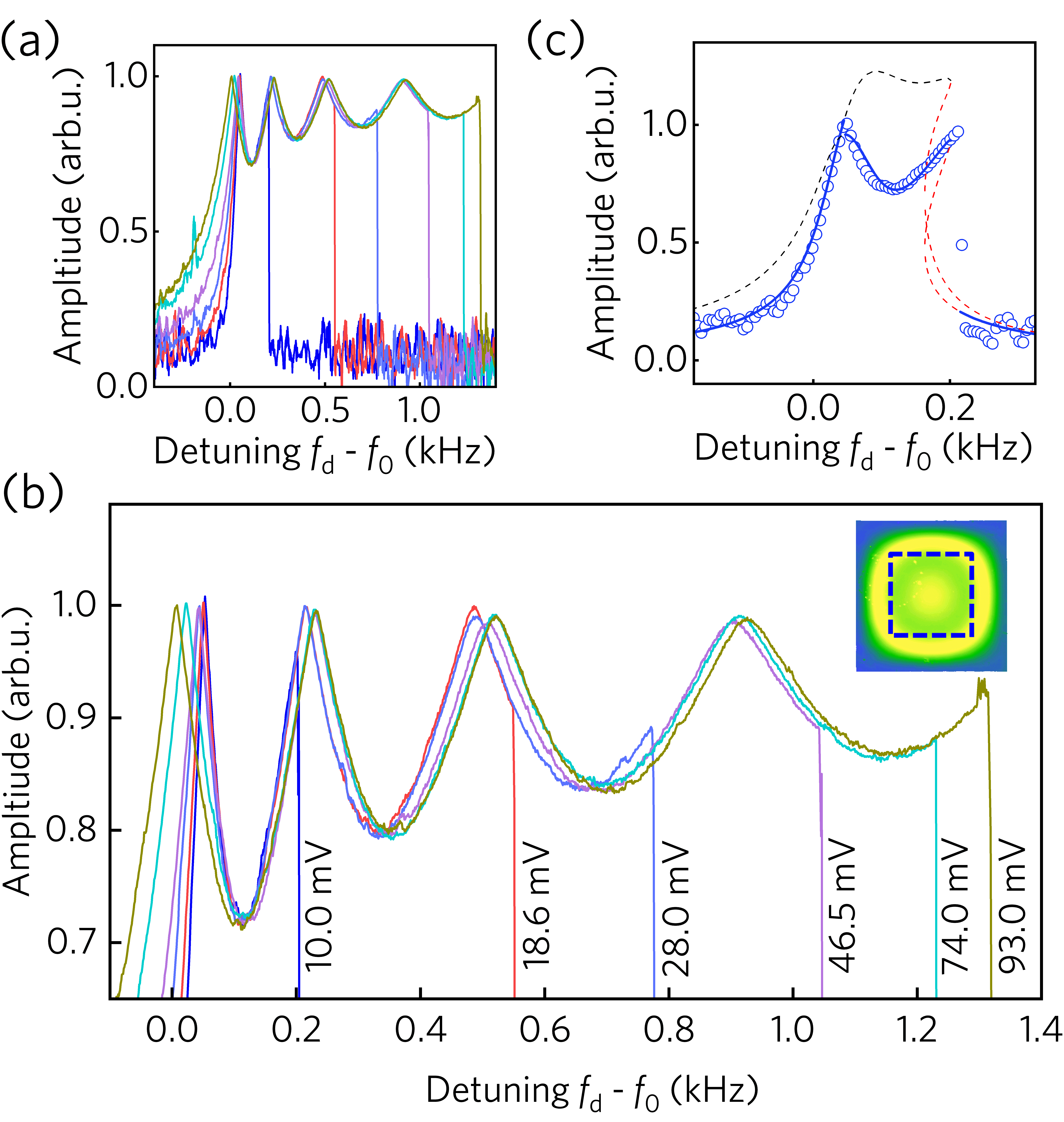}
  \caption{Resonance curves measured in a localized area. (a) Normalized averaged amplitude responses of the central area of the membrane (blue frame in inset deflection pattern), measured by IWLI curve for different $\textit{V}_{\textrm{exc}}$ from 10 mV up to 93 mV. (b) Zoom into the maximum region. 
  (c) Experimental curve (symbols) for $\textit{V}_{\textrm{exc}}$ from 10 mV and fit (line) to the 1:2 model.}
  \label{fig:image04}
\end{figure}

The analysis of the development of the spatial overtones is completed by studying the deflection amplitude in a small area. Figure~\ref{fig:image04}(a) shows examples of response curves integrated over a square with size 80 x 80 $\mu$m in the central area (see dashed line in inset) of the membrane for increasing $\textit{V}_{\textrm{exc}}$.
A zoom into the maxima is given in panel (b). It reveals a serious of maxima and minima, before the amplitude eventually breaks down. The number of observable maxima increases with $\textit{V}_{\textrm{exc}}$ in agreement with the increasing plateau length when integrating over the whole membrane area. 
For small $\textit{V}_{\textrm{exc}}$ = 10 mV only the first maximum  and minimum are accessible, for $\textit{V}_{\textrm{exc}}$ = 100 mV four maxima and minima are obtained. 
We interpret this observation as follows:   
%
Upon increasing $f_{\textrm{d}}$ the diameter of the circular shaped overtones increase. The first maximum signals that $u_1$ reaches the edges of the integration window, thereby reducing the averaged amplitude. 
When further increasing $f_{\textrm{d}}$, more overtones are excited, increasing the amplitude and subsequently leave the central area, thereby reducing the averaged amplitude again.
%
%
%
As an example we fit the curve recorded for $\textit{V}_{\textrm{exc}}$ = 10 mV with our theoretical model, comprising two tones,  
$\textit{q}_\textrm{1}$ and $\textit{q}_\textrm{2}$.
To account for the fact that upon increasing $\textit{f}_{\textrm{d}}$, the radius of the ring increases and partially leaves the integration area, we use a linearly decreasing weight factor for the mode $\textit{q}_\textrm{1}$. 
The solid line in Fig. \ref{fig:image04}(c) is the result of the fitting with the theoretical model which describes the experimental observation qualitatively correctly. The individual contributions and the total fitted amplitude and further details are given in the SM. 
This concept can straightforwardly be extended to the situation with three and more modes. However, the number of fitting parameters increases rapidly.
Summarizing, the oscillating amplitude in the central area of the membrane can be interpreted as the gradual transfer of vibrational energy from $q_1$ to higher overtones $q_n$. 
In conclusion, by combination of two complementary experimental approaches and the nonlinear $1:n$ coupling model we revealed a novel deflection state caused by a nonlinear coupling mechanism between several overtones of a single flexural mode of a membrane oscillator. 
This deflection state is hallmarked by a spatial modulation pattern, in which different parts of the membrane oscillate at different frequencies. 
For rectangular membranes the spatial modulation state can be distinguished from the linear deflection pattern by a distinctly different symmetry.
The coupling mechanism is mediated by nonlinear interaction and activated by strong driving. 
%
%
%
%
\setcounter{secnumdepth}{-1}
\section{\label{sec:level1} Acknowledgments}
The authors thank R. Waitz for help in sample fabrication. 
We are indebted to W. Belzig, T. Dekorsy, M. Dykman, V. Gusev, M. Hettich, P. Leiderer, S. Shaw, I. Wilson-Rae, R. S. Edwards, L. Q. Zhou and 
the SFB767 Nanomechanics Discussion Group for fruitful discussion and comments about the work. 
The authors gratefully acknowledge financial support from the China Scholarship Council, 
the European Union's Horizon 2020 program for Research and Innovation under 
grant agreement No. 732894 (FET Proactive HOT) and the Deutsche Forschungsgemeinschaft 
via the Collaborative Research Center SFB767.

\bibliographystyle{apsrev4-1}

\end{document}



\title{Supplementary material for "Spatial modulation of nonlinear flexural vibrations of membrane resonators"}

\author{Fan Yang}
\author{Felix Rochau}
\author{Jana Huber}
\author{Alexandre Brieussel}
\author{Gianluca Rastelli}
\author{Eva M. Weig}
\author{Elke Scheer}
 \email{elke.scheer@uni-konstanz.de}

\affiliation{Department of Physics, University of Konstanz, 78457 Konstanz, Germany}

\date{\today}

\maketitle

%
%
%
%
\section{\label{sec:level1} Sample Fabrication and measurement principles}
\subsection{\label{sec:level2} Fabrication}
\begin{figure*}[t]
   \includegraphics[width=\textwidth]{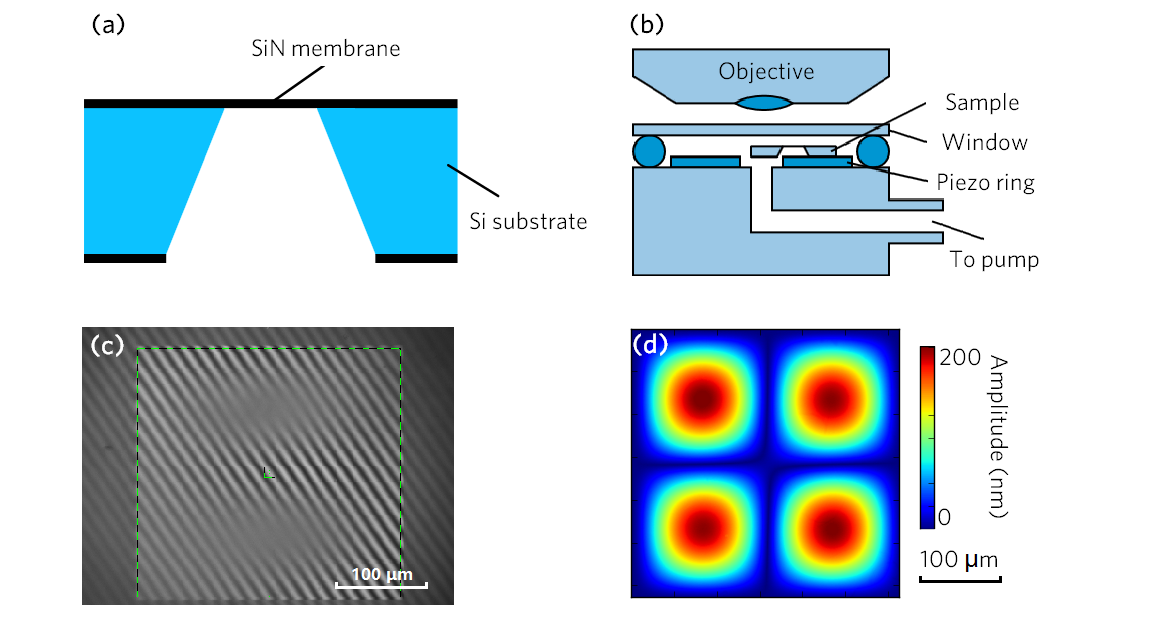}
   \caption[width=\textwidth]{Siliconnitride membrane sample and experimental setup of IWLI. (a) Cross section of a free-standing SiN membrane supported by a 0.5 mm thick silicon substrate. (b) Sketch of the IWLI measurement setup showing the membrane chip, the piezo ring, the vacuum chamber and the objective. (c) Camera view of a 298 $\times$ 296 $\mu$m$^{2}$ membrane under the imaging interferometer objective using continuous light. The measurement area is indicated by the green frame. Here the membrane is in the (1,2) mode, as indicated by the blurring of the almost diagonal interference lines. (d) False color image of an amplitude profile of the membrane recorded at an excitation frequency of 909 kHz, corresponding to the (2,2) mode. }
   \label{fig:image06}
 \end{figure*}

The siliconnitride (SiN) membranes are fabricated from a 0.5 mm thick commercial (100) silicon wafer, both sides of which are coated with $\mathrm{\sim}$ 500 nm thick Low Pressure Chemical Vapor Deposition (LPCVD) SiN. The membrane is fabricated on the front layer, and the backside layer serves as an etch mask. Laser ablation is used to open the etch mask with a typical size of 1 $\mathrm{\times}$ 1 mm$^{2}$. Using anisotropic etching in aqueous potassium hydroxide (KOH), a hole is etched through the openings of the mask. After about twenty hours the KOH solution reaches the topside layer and a membrane is formed, supported by a massive silicon frame. The cross section of a free-standing SiN membrane supported by the silicon substrate is presented in Fig.~S\ref{fig:image06}(a). In this work, a 478 nm thick and 413.5 $\mathrm{\times}$ 393.5 $\mu$m$^{2}$ lateral size membrane is employed. The chip carrying the membrane is glued using 2-component adhesive with contact points (5 mm diameter) at each corner of the substrate to a piezo ring with 20 mm diameter and 5 mm thickness.
\subsection{\label{sec:level3} Imaging white light interferometry (IWLI)}
The sample is placed in a vacuum chamber held at room temperature, connected to a pressure controller, see Fig.~S\ref{fig:image11}(b), to ensure full control over the pressure of the surrounding atmosphere in a pressure range from \textit{p} = 0.001 mbar to atmospheric pressure. The measurements discussed in the present work have been performed at 2 $\times$ 10$^{-2}$ mbar. The surface of the membrane is observed by an imaging interferometer using different light sources, described in detail in Ref. \cite{petitgrand20013d}. The excitation voltage is applied using a sinusoidal function generator the phase of which can be locked to the stroboscopic light of the imaging white light interferometer.\\
\indent The observed interference pattern represents the surface profile of the sample as exemplified in Fig.~S\ref{fig:image06}(c). The interference fringes can be used to quantitatively measure the vibrational amplitude. Figure~S\ref{fig:image06}(d) shows a measurement example for the (2,2) mode. The vibrational deflection pattern can be measured by stroboscopic illumination and by continuous illumination. The deflection pattern measured by stroboscopic light contains vibrational phase information but is limited to the specific driving frequency with a locked phase. The continuous light can be applied without any phase lock-in, and thus provides deflection amplitude patterns. In this work, we utilized continuous light for recording the mode deflection patterns, stroboscopic light was utilized to measure the image shown in Fig.~S\ref{fig:image06}(d).

\subsection{\label{sec:level3}  Michelson interferometry (MI)}
A MI is utilized to measure the sweep-up and ring-down processes of the deflection at a certain point on the membrane area. The sample mounting on the piezo ring is the same as for the IWLI, and the sample is also paced in a vacuum chamber at room temperature. However, the instrument is located in another lab with slightly different temperature. because of the sensitivity of the eigenfrequencies to temperature, small deviations in the absolute values of the eigenfrequencies in the range of a few hundred Hz may occur. When indicating absolute frequency values we use here the ones detected with the MI.  A HeNe laser ($\lambda$ = 633 nm) is focused on a spot (\textit{d} $\approx$ 1 $\mu$m) on the free-standing membrane which is positioned using an \textit{xyz} piezo-positioning stage. The vibrations of the oscillating membrane modulate the reflected light, which is interfered with a phase reference (stabilized with a control bandwidth of 48~kHz). The detected power of the interference signal is proportional to the amplitude squared for deflections much smaller than the optical wavelength. For strong excitation voltages, this limit is hard to obey. At most positions the deflection amplitude is then much larger than a quarter of the laser wavelength used in the MI. To avoid ambiguities in the data interpretation we focus the laser on a point close to one edge of the membrane, where the amplitude remains well within one fringe of the MI signal. Frequency response spectra are measured using a fast lock-in amplifier with a bin width of 1~kHz. For the frequency-resolved ring-down measurements, the oscillations are recorded with an oscilloscope with a sampling rate of 5~MS/s. Then an FFT is performed on sets of 2000 sample points each. Integration around a particular oscillation frequency yields the separated energy decay traces shown in Figs.~S\ref{fig:image10}(b) and S\ref{fig:image11}(b) below.

%
%
%
%
\section{\label{sec:level4} Characterization of mechanical properties}

\begin{table*}
  \caption{Experimental and theoretical values of different flexural modes. Measured eigenfrequencies of different flexural modes (\textit{m,n}) in kHz. The frequency values have been measured by MI (underlined) and IWLI. Calculated eigenfrequencies of different flexural modes (\textit{m},\textit{n}) in kHz. }
  \label{tab:table01}
  \begin{ruledtabular}
  \begin{tabular}{ccccccccccc} \hline
   & \multicolumn{5}{c}{Experimental} & \multicolumn{5}{c}{Theoretical} \\ \hline
  Mode & 1 (\textit{n}) & 2 & 3 & 4 & 5 & 1 & 2 & 3 & 4 & 5 \\ \hline
  1 (\textit{m}) & 321 & 500 & \underline{712} & 927 & 1174 & 325 & 507 & 714 & 930 & 1149 \\ \hline
  2 & 517 & 646 & 818 & 1018 & 1233 & 520 & 649 & 821 & 1014 & 1218 \\ \hline
  3 & \underline{747} & 837 & \underline{983} &  & 1326 & 738 & 834 & 974 & 1141 & 1326 \\ \hline
  4 & \underline{979} & \underline{1052} &  &  &  & 963 & 1039 & 1154 & 1299 & 1463 \\ \hline
  5 & 1219 &  & 1351 &  & 1640 & 1192 & 1254 & 1351 & 1477 & 1623 \\ \hline
  \end{tabular}
  \end{ruledtabular}
\end{table*}

 By applying our customized VICL (Vibrometery In Continuous Light) \cite{yang2017quantitative} and APSStro (Automated Phase-Shifting and Surface Measurement in Stroboscopic Light) measurement methods \cite{zhang2015vibrational, waitz2012mode} dispersion relations of the bending waves of the membrane are measured using IWLI. From this data, Young's modulus \textit{E} and the residual stress \textit{$\sigma$} are determined by fitting. The resonance frequencies as well as the mechanical \textit{Q} factors of different vibrational modes of this SiN membrane are quantitatively determined. For the (1,1) mode we find \textit{f}$_{(1,1)~}$ = 323.5 kHz, \textit{Q}$_{(1,1)}$ = 2 $\times$ 10$^{4}$, \textit{E} = 240.2 GPa, \textit{$\sigma$} = 0.1261 GPa. Note that temperature drifts induce frequency shifts in the order of 500 Hz/K. The temperature in the lab is stabilized with a precision of $\pm$ 1K. The IWLI and the MI are located in two different labs with slightly different average temperature. Thus, the absolute values of the eigenfrequencies may vary about $\pm$ 2.5 kHz from set-up to set-up. The absolute values indicated in the main text and in this Appendix correspond to the ones measured with MI. The full width at half maximum (FWHM) of the (1,1) mode in the linear response regime is about 50 Hz. Table 1 gives an overview over the expected eigenfrequencies of the membrane, measured by MI and IWLI, and calculated using the formula:$\mathrm{\ }f_{m,n}\cong \sqrt{{({\sigma }_{xx}{m^2}/{L^2_{\mathrm{w}}}+{\sigma }_{yy}{n^2}/{L^2_{\mathrm{h}}})}/{(4\rho )}}$. Here the mass density $\rho$ = 3.18 $\mathrm{\times}$ 10$^{3}$ kg/m$^{3}$, \textit{L}${}_{w}$ = 413.5 $\mu$m and \textit{L}$_{h}$ = 393.5 $\mu$m are the edge lengths of the membrane, and \textit{m,n} denote the integer mode indices representing the number of antinodes. The observed temperature dependence of the eigenfrequencies is attributed to temperature-dependent stress tensor components \textit{$\sigma$}${}_{xx}$ and \textit{$\sigma$}${}_{yy}$ along \textit{x} and \textit{y} axis, respectively. The values given in the table have been calculated using the values \textit{$\sigma$}$_{xx}$ = 0.110 GPa and \textit{$\sigma$}$_{yy}$ = 0.108 GPa determined by our spatially resolved measurement method described in Ref. \cite{waitz2015spatially}.
 %
 %
 %
 %
 \begin{figure*}[t]
   \includegraphics[width=\textwidth]{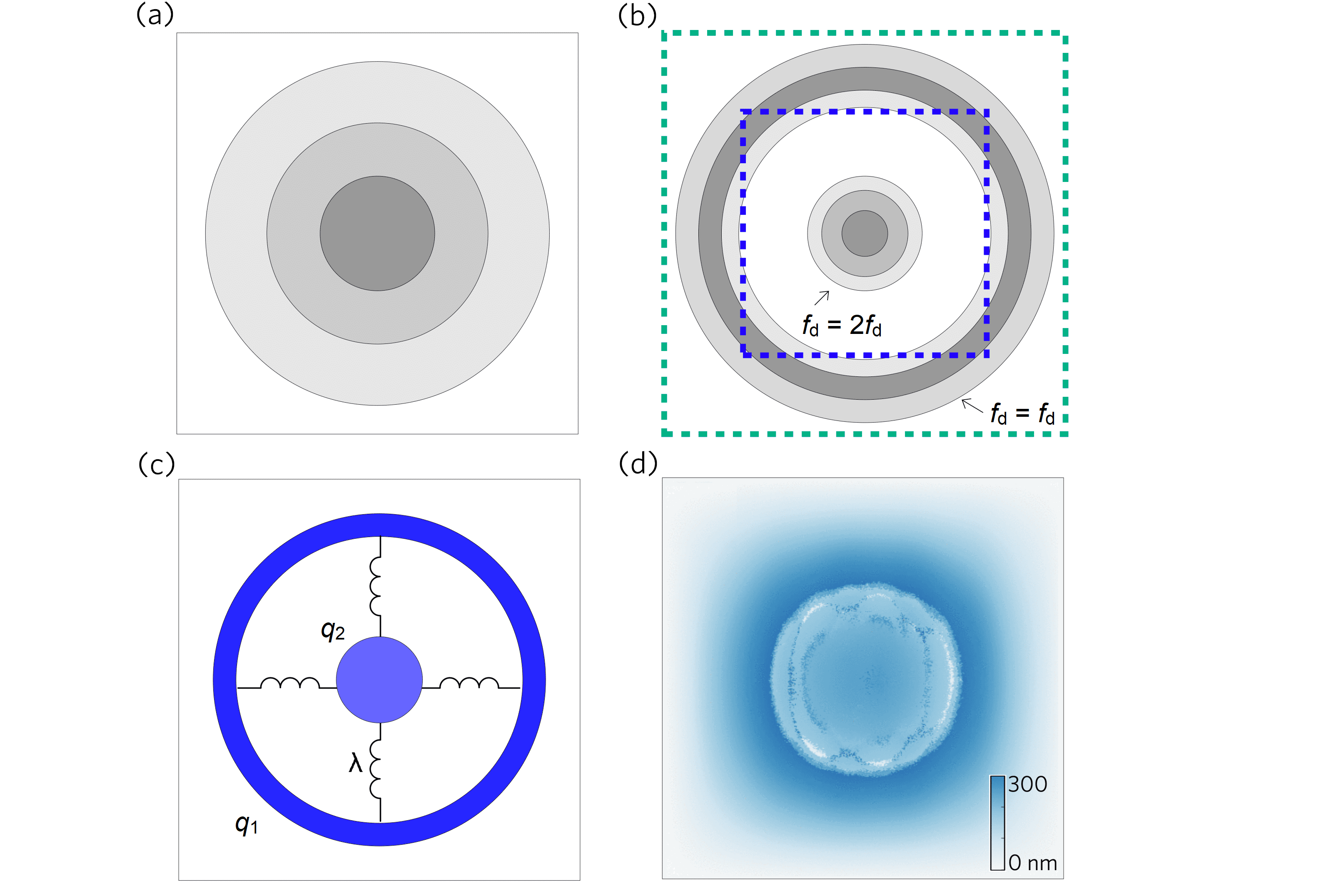}
   \caption[width=\textwidth]{Spatial modulation model. (a) The profile of the oscillation for the (1,1) mode of the membrane has circular symmetry with eigenfrequency $f_\textrm{(1,1)}$. The corresponding deflection pattern schematically depicted in Fig.~\ref{fig:image02}(b) in the main text shows a single drum head shape. (b) In the spatial modulation regime, upon increasing detuning, the membrane starts to oscillate at the center with frequency 2$f_\textrm{d}$, whereas the outer ring oscillates at frequency $f_\textrm{d}$. The corresponding profile is sketched as Fig.~\ref{fig:image02}(b) in the main text. (c) Phenomenological model of the membrane in the spatial modulation regime. The outer ring is assumed as an oscillator of eigenfrequency $f_\textrm{1}$ = $f_\textrm{(1,1)}$ and amplitude $q_\textrm{1}$ and the inner disc is assumed as an oscillator of frequency 2$f_\textrm{1}$ and amplitude $q_\textrm{2}$. The two oscillators are nonlinearly coupled with potential $V_\textrm{int} = \frac{1}{2} \lambda q^{2}_\textrm{1} q_\textrm{2}$. (d) Stroboscopic image recorded at $f_\textrm{d}$ = $f_\textrm{3}$ (see Fig.~\ref{fig:image02}(b) of the main text). The central part oscillates at different frequency than the ring which oscillates at $f_\textrm{(1,1)}$ and thus the phase of the drum head motion cannot be locked with the value of the  ring. Therefore only the deflection amplitude of the outer ring is imaged correctly, while the central part appears blurred. }
   \label{fig:image07}
 \end{figure*}

%
%
%
%
%
%
\section{\label{sec:level5} Phenomenological model}
The theoretical description of the observed complex spatial deflection profiles is a challenging task. A priori, one must go back to the governing elastic equation of the system for the membrane profile $u\left(\overrightarrow{r},t\right)$ and exploit, e.g., numerical simulations based on the decomposition of finite elements. However, we observed that for certain driving frequencies and excitations the membrane is vibrating at different frequencies in distinct spatial sectors. Our aim is to develop an analytical model for this situation. To do so we employ a distributed model, assuming that the membrane is composed by an inner disc coupled to outer rings, see Fig.~S\ref{fig:image07}. In that case the membrane profile is described by Eq.~\eqref{eq01} of the main text. A similar approach was used in silicon ring resonators \cite{bijari2012nonlinear}. The inner disc and the outer rings play the role of effective modes. We assume that these modes are nonlinearly coupled such that the motion of the low frequency mode, the outermost ring $u_\mathrm{1}$, actuates the higher frequency modes, i.e. the inner rings and central disc, $n \geq 2$. 

\begin{figure}[htbp]
   \includegraphics[width=\linewidth]{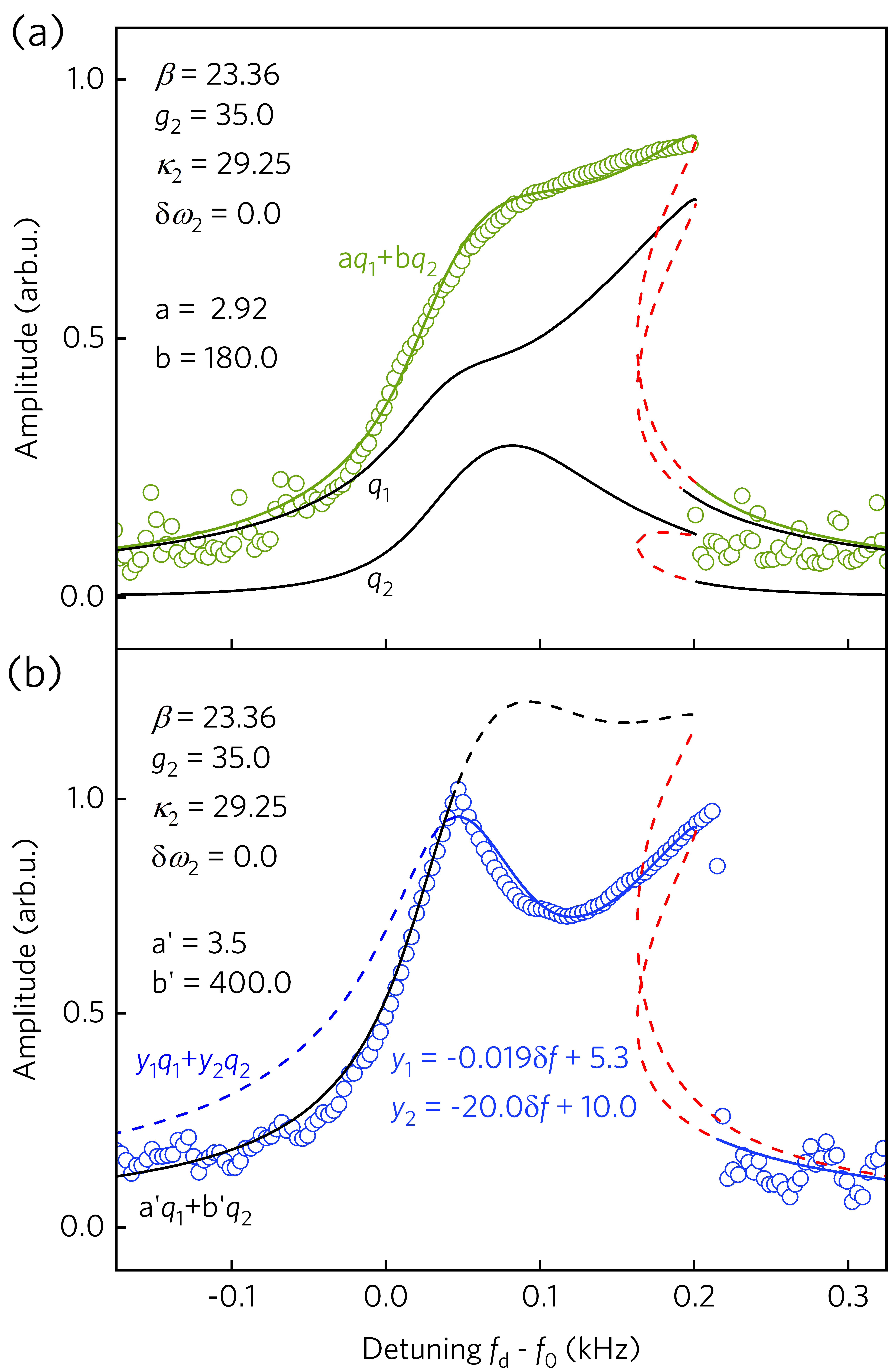}
   \caption[width=\linewidth]{Decomposition of the modeling of spatial modulation. The 1:2 internal resonance of nonlinear coupling discussed above is used for the theoretical modeling. A set of suitable parameters is applied for the central area and full area (see Fig.~S\ref{fig:image07}) of the membrane, IWLI measured resonance curves (also shown in Fig.~\ref{fig:image02} in the main text). (a) The nonlinear response measured over the full area of the membrane is plotted as green dots with the theoretical curve a$q_1$ + b$q_2$. The individual theoretical modelling of $q_1$ and $q_2$ are plotted with black solid curves (stable solutions), the red dashed lines denote unstable solutions. (b) The nonlinear resonance response measured in the central area of the membrane is plotted as blue dots with the theoretical modelling curves $y_1q_1$ + $y_2q_2$ (blue) and a'$q_1$ + b'$q_2$ (black). The fitting parts of the two modelling curves are plotted as solid lines; the ineffective parts are plotted as dashed lines.}
   \label{fig:image08}
 \end{figure}

The starting model is given by the nonlinear interaction between the fundamental mode $n=1$,
with angular eigenfrequency $\omega_1\sim \omega_\textrm{d}$, and other modes $\omega_n = n \omega_1$.
The dynamical equations read
%
%
%
%
%
\begin{align}
\ddot{q}_1(t)
& =
- \omega_{1}^2  q_1(t)
- 2 \Gamma_1 \dot{q}_1(t)
- \gamma q^3_1(t)
+ F_{\mathrm{d}} \cos \left( \omega_d t \right)  \nonumber \\
& \phantom{=}  - \sum_{n\geq 2} \, n \, \lambda_n \,  q_1^{n-1} \,  q_n   \, ,
\label{eq:n_model-1} \\
\ddot{q}_n(t)
& =
- \omega^2_n  q_n(t)
- 2 \Gamma_n \dot{q}_n (t)
- \lambda_n \, q^n_1  \quad (n \geq 2)\, ,
\label{eq:n_model-2}
\end{align}
in which  we have  neglected the self nonlinearity  of the higher modes since
we expect that they have a lower oscillation amplitude as their motion is activated
by the weak nonlinear interaction with the fundamental
mode given by the potential   $V = \sum_{n\geq 2} \lambda_n q_1^n q_n$.
%
%
%
Using the canonical transformations
%
%
%
%
%
\begin{align}
\label{eq:n_model-3}
u_n (t) 	&=   e^{-i \, n \,  \omega_{\mathrm{d}} t} \left[  q_n (t) - i  \dot{q}_n(t) / (n \omega_d)  \right] /2  \, , \\
u^*_n (t) 	&=   e^{i \, n \,  \omega_{\mathrm{d}} t} \left[  q_n (t) + i  \dot{q}_n(t) / (n \omega_d)  \right] /2   \, ,
\end{align}
and applying the rotating wave approximation (RWA), we arrive at the following equations
%
%
%
%
%
%
\begin{align}
\dot{u}_1(t)
= &
-\left( \Gamma_1 +i\delta\omega
-i\frac{3\gamma }{ 2\omega _{\mathrm{d}} } {\left| u_1 (t) \right|}^2
\right)
u_1(t)
-\frac{iF}{4{\omega }_{\mathrm{d}}}  \nonumber \\
&
+ \sum_{n \geq 2}
\frac{i n \lambda_n }{2 \omega_{\mathrm{d}}} \,   u_n(t)  \,  {\left( u_1^{*}(t)\right)}^{n-1} \, ,
\label{eq:n_model-5}
\\
\dot{u}_n(t)=
&
-\left( \Gamma_n +i n \delta\omega \right) u_n (t)
+
\frac{i\lambda_n }{ 2 n \omega _{\mathrm{d}} } u^{n}_1(t)  \quad  (n \geq 2)
\label{eq:n_model-6}
\, .
\end{align}
with $\delta\omega =  \omega_d - \omega_n$.
The  variables $u_n$ describe the oscillation amplitudes in the rotating frame at the driving frequency $\omega_d$.
Hence in the steady state, $\dot{u}_n(t)=0$, we obtain the stationary solution $\bar{u}_n$ given by the equations
%
%
%
%
%
%
\begin{equation}
0
\!\!=\!\!
\left[
\!
\Gamma_1 +i\delta\omega
-i\frac{3\gamma }{ 2\omega _{\mathrm{d}} } {\left| \bar{u}_1  \right|}^2
\!
\right]
\!
\bar{u}_1
+
\!
\frac{iF}{4{\omega }_{\mathrm{d}}}
\!
-
\! \sum_{n \geq 2}
\frac{i n \lambda_n }{2 \omega_{\mathrm{d}}} \,   \bar{u}_n  \,  {\left( \bar{u}_1^{*}\right)}^{n-1}
\!\!\!\! ,
\label{eq:n_model-7}
\end{equation}
and
%
%
%
%
%
%
\begin{equation}
0 =
\left( \Gamma_n +i n \delta\omega \right) \bar{u}_n
-
\frac{i\lambda_n }{ 2 n \omega _{\mathrm{d}} } \bar{u}^{n}_1 \quad  (n \geq 2)
\label{eq:n_model-8}
\, .
\end{equation}
We use now the following scaling for the amplitudes
%
%
%
%
%
%
\begin{equation}
\bar{u}_n =   \sqrt{\frac{2\omega_d \Gamma_1}{3\gamma}}  \, z_n  \,
\label{eq:n_model-9}
\end{equation}
and set the scaled parameters as the scaled detuning $\Omega = \delta \omega /  \Gamma_1$,  the scaled damping
$\kappa_n = \Gamma_n/\Gamma_1$, and
%
%
%
%
%
%
\begin{equation}
\beta = F \sqrt{ \frac{3\gamma }{32 \omega_d^3 \Gamma_1^3} }
\, , \quad
g_n =  \frac{\lambda^2_n}{4 \omega_d^2  \Gamma_1^2} {\left(  \frac{2\omega_d \Gamma_1}{3\gamma}  \right)}^{n-1} \,
\label{eq:n_model-10}
\end{equation}
with  $\beta$  the scaled force provided by the external drive.
From the Eqs.~S(\ref{eq:n_model-5}) one can obtain a closed equation for the amplitude of the mode $\bar{u}_{n=1}$. The
equation for the stationary amplitude $z_1$ reads
%
%
%
%
%
%
\begin{equation}
0
=
\left[
1 +i \Omega
-i {\left| z_1  \right|}^2
+ \sum_{n \geq 2}   \frac{g_n {\left| z_1 \right|}^{2(n-1)}  }{\kappa_n +i  n \delta\omega  }
\right]
z_1
+ i \beta
\, ,
\label{eq:n_model-11}
\end{equation}
whereas  the scaled oscillation amplitudes  for $n\geq 2$ are
%
%
%
%
%
%
%
\begin{equation}
{\left| z_n \right|}^2 =  \frac{ \left( g_n/n^2 \right) }{\kappa_n^2 +  n^2 \Omega^2}
{\left| z_1  \right|}^{2n}
\, .
\label{eq:n_model-12}
\end{equation}
\\\indent Based on the previous coupling model, we explicitly solve the special case involving only two modes.
%
%
In this case, $\omega_1$ and  $\omega_2$ are the angular eigenfrequencies of the ring mode and the center (inner) mode,
respectively, and $\lambda_{12}=\lambda$ is the coupling strength.
%
%
%
%
This simplification leads to Eq.~(3) and Eq.~(4) in the main text which
are examples of the internal coupling resonance of the type $1:n$
(with potential energy $V_n = \lambda_n q^n_1 q_2$
and $\omega_n  \approx   n \omega_1$)
that has received recently great attention - in particular for the case 1:3 -
in experiments reported in Refs.~\cite{guttinger2017energy, antonio2012frequency, chen2017direct}
and in the theoretical studies Refs.~\cite{shoshani2017anomalous, mangussi2016internal, shoshani2017three}.
Equation~S\eqref{eq:n_model-11} then reads:

%
%
%
\begin{equation}
\left[
1 +i \Omega
-
\left(
i
-
  \frac{g_2   }{\kappa_2 +i  2 \delta\omega  }
\right) {\left| z_1  \right|}^2
\right]
z_1
= -  i \beta
\, ,
\label{eq:n_model-13}
\end{equation}
which shows that  the case  $n=2$ is special, since we obtain an equation similar to the Duffing model.
The amplitude square of the second mode is given by
%
%
%
%
%
%
\begin{equation}
{\left| z_2 \right|}^2 =  \frac{ \left(g_2/4\right) }{\kappa_2^2 + 4 \Omega^2}
{\left| z_1  \right|}^{2}
\, .
\label{eq:n_model-14}
\end{equation}
In the spatial modulation regime with one inner drum head mode and one outer ring, the
membrane vibration is described by
%
%
%
%
%
%
\begin{equation}
h\left( \vec{r},t \right) \approx
h_1(\vec{r})  \,  \bar{u}_1 e^{i \omega_d t}   \,  + \,   h_2(\vec{r})  \,  \bar{u}_2 e^{2 i \omega_d t} \
+ \mbox{c.c.}  \, ,
\label{eq:n_model-15}
\end{equation}
with $h_1(\vec{r})$ the profile function of the ring and
$h_2(\vec{r}))$ the profil function of the center, the latter similar to the profile
of the mode $(1,1)$.
On the other hand, the experimentally measured quantity by IWLI represents an average over the membrane,  namely
%
%
%
%
%
%
\begin{equation}
A_{exp}^2 \propto \frac{1}{T} \int^{T}_{0}\!\!\!\!\! dt  \iint^{ \frac{L_e}{2} }_{-\frac{L_e}{2}  } \!\!\!\!\!  dx dy \,\,    {\left| h\left( \vec{r}, t \right)   \right|}^2
\label{eq:n_model-16}
\end{equation}
with $T=2\pi/\omega_d$, $\vec{r}\equiv(x,y)$ and $L_e$ the size of the observed square on the membrane.
Since neither the two shape functions $h_1(\vec{r})$ and $h_2(\vec{r})$, nor the
phase difference between the two parts oscillating at $\omega_d$ and $2 \omega_d$ can be measured,  we introduce two weight factors to fit the phenomenological model to the experimental data. To do so we use the following simple ansatz
%
%
%
%
%
%
\begin{equation}
\bar{A} \propto \left( a \left| z_1 \right| + b \left| z_2 \right| \right)  \,.
\label{eq:n_model-17}
\end{equation}
Note that in the single mode regime $h\left( \vec{r},t \right) \approx
h_1(\vec{r})  \,  \bar{u}_1 e^{i \omega_d t}   + \mbox{c.c}$, the proportional scaling
$\bar{A} \propto |z_1|$  indeed holds.
We now turn to exploit the nonlinear $1:n$ model in practice.
Equation~S(\ref{eq:n_model-17}), with constant weighting factors $a$ and $b$,
represents a very first approximation to extract some quantitative comparison with the experiment.
Solving Eqs.~S(\ref{eq:n_model-13}) and S(\ref{eq:n_model-14}), we use Eq.~S(\ref{eq:n_model-17})
to fit the experimental amplitude averaged over the full area of the membrane $L_e =  L$, see Fig. S\ref{fig:image07}(b).
The results are reported in Fig.~S\ref{fig:image08}(a) for $V_{\textrm{exc}}=10~\mathrm{mV}$ (as in the inset of the Fig.~3 of the main text).
The two black solid lines are the individual amplitudes $|z_1|$ and $|z_2|$ with
the dashed red lines for the unstable parts of the solutions.
The solid light blue line is Eq.~S(\ref{eq:n_model-17}) with
the fitted parameters $\beta = 23.36$,  $\kappa_2 =29.25$  and $\textit{g}_2 = 35.0$
and weighting factors $a = 2.92$ and $b = 180.0$.
In view of the simplicity of the model the, curve reproduces the experimental data well.

Effectively, as described in the main text, the vibrational shape of the membrane
changes as varying the detuning $\delta\omega$ even at fixed drive $V_{\textrm{exc}}$.
This is particular relevant when we measure the average amplitude
in a square over the central part with smaller size  $L_e = \frac{1}{16}L$
of the entire membrane: since the diameter of the ring increases with the detuning to accommodate the higher modes, the ring eventually leaves the integration square $L_e^2$. In this situation the average amplitude decreases drastically, resulting in a pronounced peak, as observed in Fig.~\ref{fig:image03} in the main text.
An example, how this can be described within the nonlinear coupling model, is shown in Fig.~S\ref{fig:image08}(b) in which we report
the measured resonance curve integrated over
the central area of the membrane
(marked as blue square frame in Fig.~S\ref{fig:image07}(b) and in the inset of Fig.~\ref{fig:image04}(b) of the main text).
The formula with fixed weighting factors is now used for frequencies below the first response maximum only, and we obtain $a= 3.5 $ and $b= 400.0$, plotted as solid black line (and continued as dashed line beyond the maximum).
When further increasing the detuning,
the ring and drum head amplitudes
partially move out of the observation frame and the average amplitudes start to decrease.
To take this into account, the weighting factors are not kept fixed, instead beyond this peak we use the fitting formula
%
%
%
%
%
\begin{equation}
\bar{A} \propto \big(  y_1(\delta\omega)  \left| z_1 \right| + y_2(\delta\omega) \left| z_2 \right| \big)  \,
\label{eq:n_model-18}
\end{equation}
with $y_1(\delta\omega) =c_0 -  c_1  \delta\omega/\Gamma_1$ and  $y_2(\delta\omega) =d_0-   d_1 \delta\omega/\Gamma_1$.
The result is the blue solid line in Fig.~S\ref{fig:image08}(b)
with fitted parameters $c_0 =5.3$, $c_1=0.019$, $d_0=10.0$ and $d_1= 20.0$.
We note that the system parameters $\kappa_2$, $\beta$ and $g_2$  are the same as those obtained from the fitting over
the full membrane, demonstrating the consistency of the phenomenological model.
 %
 %
 %
 %
%
%
 \begin{figure}[htbp]
   \includegraphics[width=\linewidth]{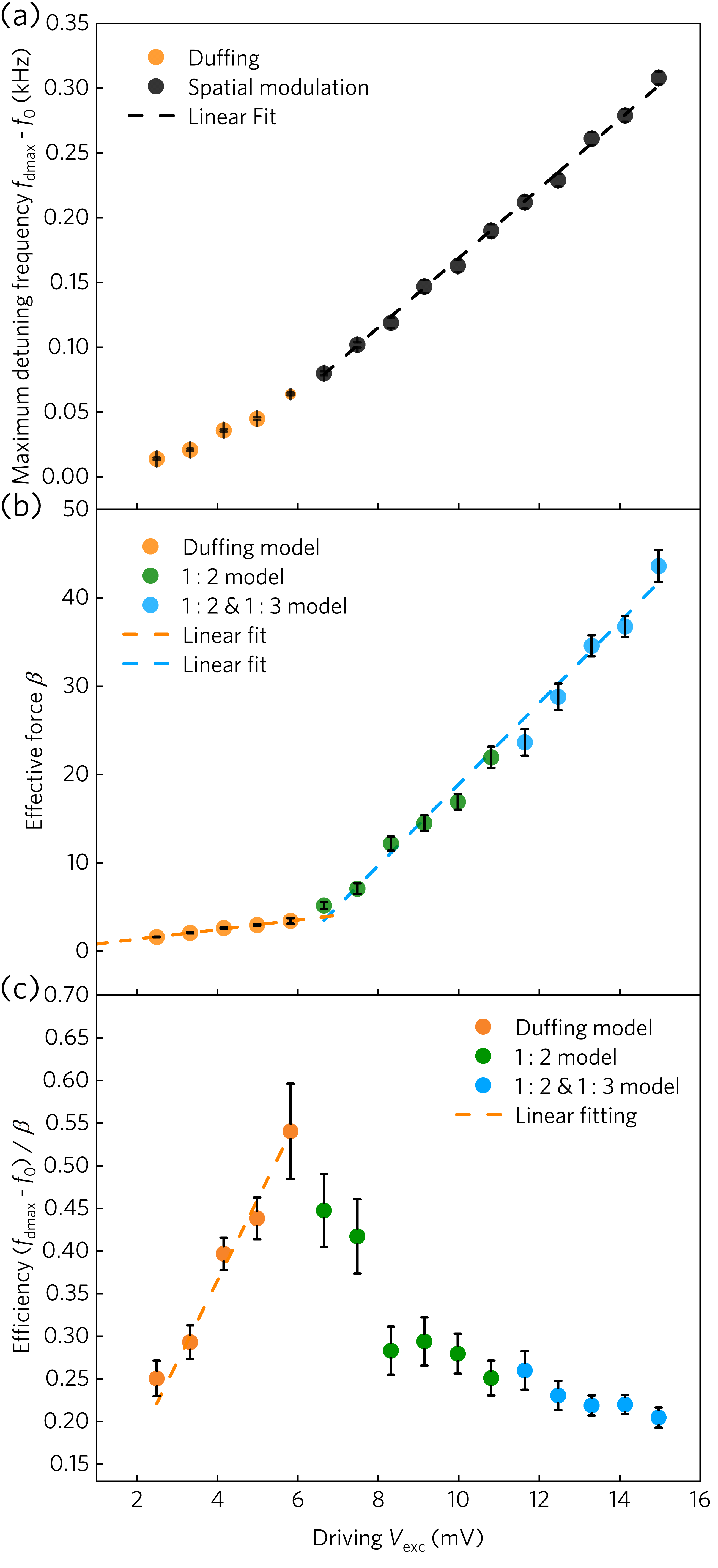}
   \caption[width=\linewidth]{Fitting parameters for the whole-membrane measurements under different excitation. (a) The maximum detuning frequency of each curve is plotted, curves in the Duffing regime are plotted as orange, and the curves in the spatial modulation regime are plotted as black. (b) Modeled effective force $\beta$ of each curve. Color codes correspond to Fig.~\ref{fig:image03} in the manuscript, the orange, green and blue represent the Duffing model, 1:2 coupling model and 1:2 $\&$ 1:3 coupling model, respectively. (c) The efficiency of excitation as defined in the text. The color code corresponds to Fig.~\ref{fig:image03} in the manuscript.}
   \label{fig:image13}
 \end{figure}
 %
 %
%
%
%
%
%
\begin{figure}[htbp]
  \includegraphics[width=\linewidth]{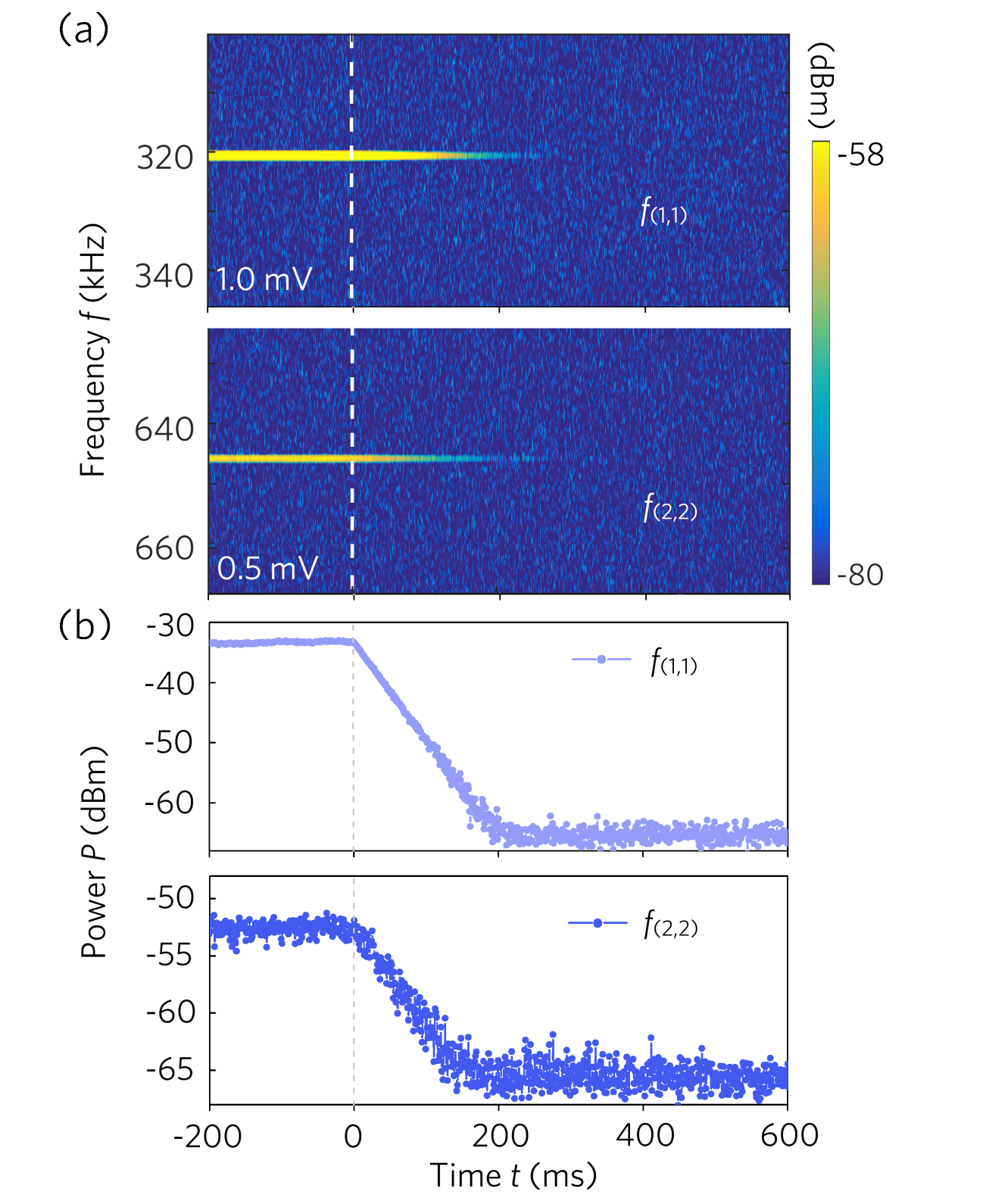}
  \caption[width=\linewidth]{ Energy decay in the linear response regime. (a) The ring-down measurements of the (1,1) and (2,2) modes are performed after weakly driving at their eigenfrequencies $f_{(1,1)}$ = 321 kHz and $f_{(2,2)}$ = 646 kHz, respectively. Panel (b) shows the time-resolved FFT power spectra obtained by integrating over a certain frequency range around the eigenfrequencies. The ring-down traces present purely exponential decays with fitted decay rate for each mode 2$\Gamma_{(1,1)}$ = 20.0 s$^{-1}$ and 2$\Gamma_{(2,2)}$ = 17.0 s$^{-1}$. }
  \label{fig:image10}
\end{figure}
We now give examples for the application of the model when more than two modes are involved.
 Figure~\ref{fig:image03} in the main manuscript shows the fitting results of the experimental curves recorded at varying $V_{\textrm{exc}}$ integrated over the entire membrane area using the $1:n$ nonlinear coupling model. The curves with intermediate drive strength ($V_{\textrm{exc}}$ = 6.7 to 10.8 mV) have been fitted with the 1:2 model. For stronger dive, the 1:2 model was not sufficient to describe the experimental findings. Motivated by the observation of the IWLI measurements showing another ring appear and the MI ring-down results also revealing the appearance of another overtone, we fitted the resonance curves with $V_{\textrm{exc}}$ = 11.6 to 15.0 mV using three modes, $n = 1,2,3$, in which the ground mode $n = 1$ is coupled to $n = 2$ and to $n = 3$ (referred to as 1:2 $\&$ 1:3 model). The best-fit parameters, the damping of the nonlinear system is $\Gamma_1 = 0.069$ $\textrm{s}^{-1}$ and the Duffing coefficient $\gamma$ amounts to $7.0 \times 10^4$ $\textrm{m}^{-2}\cdot\textrm{s}^{-2}$. \\
 \indent The maximum detuning frequency of each curve, determined as the frequency at which the amplitude breaks down, depends on the $V_\textrm{exc}$, as shown in Fig. S\ref{fig:image13}(a). In the Duffing regime, a slight increase of $f_\textrm{dmax}$ vs $V_\textrm{exc}$ is observed that turns into a linear behavior with larger slope in the spatial modulation regime (shown as black dots). This increase in slope is consistent with the shallow increase of the resonance amplitude. Since the area under the resonance curve is a measure for the dissipated energy, the break-down frequency has to increase markedly. The modeled effective force $\beta$ of each curve is shown in Fig. S\ref{fig:image13}(b), the colors of the symbols correspond to the color code of the curves in Fig. \ref{fig:image03} in the main text. 
$\beta$ increases with $V_\textrm{exc}$, with different slopes in the Duffing regime and the spatial modulation regime. There is no change in slope at the cross-over from the 1:2 to the 1:2 $\&$ 1:3 model.  The ratio $\left(f_\textrm{dmax} - f_0\right) / \beta$ is a measure for the efficiency with which the excitation is translated into deflection and is plotted in Fig.~S\ref{fig:image13}(c). In the Duffing regime the efficiency increases roughly linearly with $V_\textrm{exc}$. It reaches a maximum at the transition to the spatial modulation regime and then decreases markedly.  This observation in agreement with the interpretation of redistribution of energy from one single fundamental mode to other overtones in the spatial modulation regime. There is no qualitative change at the cross-over between the 1:2 and the 1:2 $\&$ 1:3 model, underlining the consistency of the model.
The $1:n$ model can also be used to describe the resonance curves recorded for large $V_{\textrm{exc}}$ and displaying multiple maxima averaged over the central area, only, thereby using weighting factor as described above. However, because of the large number of necessary fitting parameters, their values are less meaningful, and we refrain from reporting them here.   
\\
\begin{figure}[htbp]
  \includegraphics[width=\linewidth]{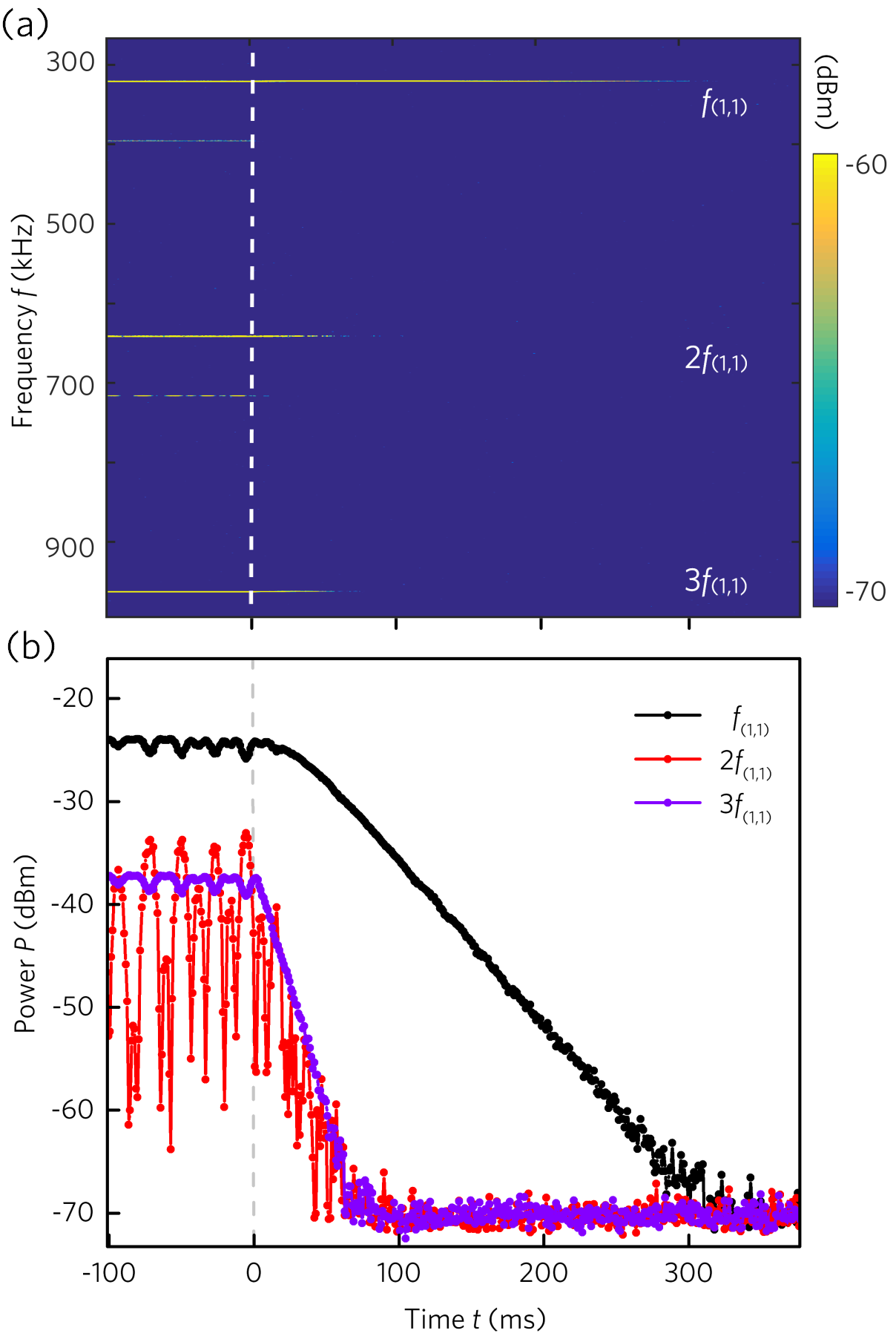}
  \caption[width=\linewidth]{ Energy decay of ultra-strongly driven (1,1) mode in the spatial modulation regime. The (1,1) mode is driven in the spatial modulation state with $V_\textrm{exc}$ = 10 mV, $f_\textrm{d}$ = 321.2 kHz slightly higher than the eigenfrequency $f_{(1,1)}$ = 321 kHz and switched off at \textit{t} = 0. (a) Power spectrum of ring-down measurement. The power trace of the (1,1) mode can be observed as well as its 2$^\textrm{nd}$ and 3$^\textrm{rd}$ overtones at 2$f_\textrm{d}$ = 642.4 kHz and 3$f_\textrm{d}$ = 963.6 kHz. Two weak and fast decaying frequency responses at 396 kHz and 716 kHz are also observed. The overtones are labelled as $f_{(1,1)}$, 2$f_{(1,1)}$ and 3$f_{(1,1)}$. (b) The time-resolved and frequency integrated power spectra of the data shown in a for different frequency ranges ($f_{(1,1)}$ black, 2$f_{(1,1)}$ red, 3$f_{(1,1)}$ blue). The amplitude of the (1,1) mode persists for around 40 ms and then decays exponentially, the 2$^\textrm{nd}$ and the 3$^\textrm{rd}$ overtone decay immediately after switching off the drive and with a larger decay rate.}
  \label{fig:image11}
\end{figure}
\section{\label{sec:level6} Ring-down measurements and energy decay}
\subsection{\label{sec:level7}  Linear response}
Figure~S\ref{fig:image10} shows the ring-downs of the (1,1) and the (2,2) mode, respectively, after driving them resonantly in the linear response regime. Both ring-downs show purely exponential decay with decay rates $2\Gamma_{(1,1)} \approx 20.0\ \mathrm{s}^{-1}$ and $2\Gamma_{(2,2)} \approx 17.0\ \mathrm{s}^{-1}$. No overtones or higher modes are excited, confirming the purely single-frequency excitation as well as the harmonic nature of the resonator.\\
\subsection{\label{sec:level8}  Nonlinear response with resonant drive in the spatial modulation regime}
Figure~S\ref{fig:image11} presents a ring-down measurement of the membrane after driving resonantly at $f_{(1,1)}$ but with ${}_{\ }$larger amplitude $V_\textrm{exc}$ = 10 mV, showing several overtones of the (1,1) mode as discussed in the main text (see discussion of Fig.~\ref{fig:image03}): Three significant frequencies are detected $f_\textrm{d}$ =~321.2 kHz, 2$f_\textrm{d}$ = 642.4 kHz and 3$f_\textrm{d}$ = 963.6 kHz. With the excitation switched off, the frequencies decrease towards the eigenfrequencies of the overtones and reach $f_{(1,1)}$ = 321.0 kHz, 2$f_{(1,1)}$ = 642.0 kHz and 3$f_{(1,1)}$ = 963.0 kHz roughly at the same time. The shift amounts to only a few hundred Hz to a couple of kHz and is therefore hard to see in Fig.~S\ref{fig:image11}, but can be clearly detected when zooming into the individual frequency ranges. Figure~ S\ref{fig:image11}(b) presents the time-resolved FFT power spectra obtained by integrating over a certain frequency range around the individual resonance frequencies. The ring-down traces present anomalous decay behavior for $f_{(1,1)}$ and its 2${}^{nd}$ overtone which reveal an oscillating faster decay. The signal powers of 2$f_{(1,1)}$ and 3$f_{(1,1)}$ are weaker than the one of $f_{(1,1)}$. The signal power of $f_{(1,1)}$only starts decaying 40 ms after the drive has been turned off. Then starts a slow anomalous decay process between 40 ms to 60 ms. The decay traces of the $f_{(1,1)}$power shows an exponential decay process 60 ms after the power off. The decay rates of the individual overtones are calculated by fitting an exponential function to the ring-down power integrated over the respective frequency range. We obtain the decay rates in the spatial modulation regime 2$\Gamma_{(1,1)}$ = 21.0 s$^{-1}$ and 2$\Gamma_{3(1,1)}$ = 64.3 s$^{-1}$. The decay rate of 2$\Gamma_{(1,1)}$ is fitted from the $f_{(1,1)}$ ring-down trace after 60~ms, and 2$\Gamma_{3(1,1)}$ is fitted from the entire ring-down trace by using an exponential decay model. The power oscillation of 2$f_{(1,1)}$can be directly seen even at the steady state with power on, the frequency is roughly 44 Hz. Remainders of this oscillation are also observable during the 2$f_{(1,1)}$ ring-down process. We argue that this extremely slow oscillation might be due to heat redistribution caused by thermoelastic processes. Further experiments are necessary to clarify the origin of this behavior that is systematically observed for 2$f_{(1,1)}$.\\

\bibliographystyle{apsrev4-1}